\documentclass[11pt,a4paper,twoside,groupcitations]{article}
\usepackage[]{latexsym,amsmath,amssymb}
\usepackage{jheppub}
\usepackage{color}
\usepackage{braket}
\usepackage{amsfonts, bbold, mathtools, physics}
\newcommand{\be}{\begin{eqnarray}}
\newcommand{\ee}{\end{eqnarray}}

\newcommand{\Det}{\text{Det}\,}
\title{Constraining the UV with the electroweak effective action}
\author[a]{Iber\^e Kuntz,}
\author[a]{Amanda Malagi}
\affiliation[a]{Departamento de F\'isica, Universidade Federal do Paran\'a,
\\
PO Box 19044, Curitiba -- PR, 81531-980, Brazil}
\emailAdd{kuntz@fisica.ufpr.br}
\emailAdd{amandamalagi@ufpr.br}

\abstract{
By considering an arbitrary bare action describing BSM physics, we use the Barvinsky--Vilkovisky resummation to obtain the most general non-local electroweak effective action at second order in the field strength. We also include the contribution of the functional measure to the effective action, which is found to modify the Higgs potential by shifting its vacuum value. The resulting effective action provides one-loop corrections to the $W$ and $Z$ boson masses, ultimately leading to the most general expression for the $\rho$ parameter at one-loop. The functional measure plays a pivotal role as it allows the parameterization of $\rho$ in inverse powers of the scale of new physics, while containing non-local form factors. The comparison of $\rho$ with the latest data leads to several constraints on the UV particle spectra of BSM models. 
}

\begin{document}
\maketitle
\flushbottom
%
%

%
%
%
%
%
%
\newpage
\section{Introduction}
\label{S:intro}

The standard model has been very successful in describing elementary particles and their interactions \cite{ParticleDataGroup:2004fcd}. Since its inception about 40 years ago \cite{Weinberg:1967tq,Salam:1968rm,Glashow:1970gm}, many experiments have been set to put it through its paces. Along these four decades, its predictions have all matched the data of accelerators, making the standard model one of the joys of modern physics.  

On the other hand, as a fundamental description of Nature, one would expect the standard model to account for the dark sector and quantum gravity, but it fails at that. Although yet quite elusive, both dark matter \cite{Workman:2022ynf} and quantum gravity \cite{Carlip:2001wq} require degrees of freedom not contained in our theory of particle physics. This suggests that physics beyond the standard model (BSM) must exist at higher energies. A more fundamental theory of particles, however, remains unknown.

The search for BSM physics had until very recently clashed with the experimental success of the standard model. Indeed, there was no significant evidence to support the existence of some more fundamental theory. In April 2022, however, the Fermilab announced their measurement of the $W$ boson mass, whose value deviated from the standard model prediction by $\sim 7\sigma$ \cite{CDF:2022hxs}. The news had been well received by a large portion of the community as a first evidence for new physics. However, such deviation also implied in a deviation from the LHC data obtained thus far. Indeed, the latest LHC's findings from March 2023 for the $W$ boson mass agreed with its past measurements and hence with the standard model \cite{ATLAS:2024erm}. At the moment, we thus face an experimental tension between the measurements performed at the LHC and at the Tevatron \cite{Amoroso:2023pey}.

While we await for the future resolution of this issue, minimal extensions of the standard model can be explored in a model-independent fashion using effective field theory. This topic has indeed received a lot of attention lately \cite{Brivio:2017vri,Isidori:2023pyp,Ellis:2021kzk,Drozd:2015rsp,Chakrabortty:2023yke}, but a full account of one-loop corrections has so far been missing. The effective action can indeed reveal non-local and non-analytical corrections that might be used to account for new physics. Moreover, the configuration-space geometry also imparts one-loop corrections via the functional measure, which must be taken into account when one deals with the Wilsonian effective theory.

The purpose of this paper is to advance in this direction by computing the most general one-loop correction, at second order in the field strength, to some electroweak's oblique observables for a theory with arbitrary fields. There are mainly two novel aspects in our calculation: i) the contribution of the most general non-local form factor at second order in the field strength and ii) the contribution of the functional measure which reflects a non-trivial configuration space. These contributions, being non-analytic in the momentum, correspond to the IR portion of some (still unknown) UV completed theory, truly corresponding to its model-independent part.
Our result allows the parameterization of physics beyond the standard model in terms of a few free parameters, namely the number of degrees of freedom, the effective couplings and the scale of new physics. The last one shows up as a direct consequence of the configuration-space geometry.

This paper is organized as follows. In Sec.~\ref{EW}, we review the Barvinsky-Vilkovisky formalism for the resummed effective action at one-loop and we use it to obtain the non-local contributions to the electroweak effective action. In Sec.~\ref{measure1}, we describe the role of the functional measure in the electroweak sector, which modifies the Higgs potential. We combine the results of Secs.~\ref{EW} and \ref{measure1} to study the Higgs mechanism and calculate corrections to the boson masses and to the $\rho$ parameter in Sec.~\ref{Higgs}. Here we also discuss the different constraints imposed by the latest data. Our conclusions are drawn in Sec.~\ref{conc}.

\section{The electroweak effective action}
\label{EW}

The 1PI effective action $\Gamma[\phi]$ is a central object in the functional formulation of quantum field theory. Albeit containing the same information as the other generating functionals, such as the partition function $Z[J]$, it provides a more transparent interpretation in terms of the mean field in place of the rather abstract source $J$. This is particularly important for the study of effective equations of motion, where $\Gamma[\phi]$ naturally generalizes the classical action $S[\phi]$ in Hamilton's principle. 

In Euclidean signature, the effective action is a solution to \cite{DeWitt:2003pm}:
\begin{equation}
	\exp{-\Gamma[\varphi]} 
	= 
	\int\mathcal{D}\Phi \, \mathcal{M}[\Phi] 
	\exp{
		-S[\Phi] 
		+ (\Phi^i-\varphi^i) \frac{\delta\Gamma[\varphi]}{\delta\varphi^i}}
  \ ,
\label{eq:pathint}
\end{equation}
where $\mathcal{D}\Phi = \prod_i \mathrm{d}\Phi^i$.
Despite its complicated nonlinear nature, when combined with the background field method Eq. \eqref{eq:pathint} allows for the application of a wide range of approximate techniques, thus providing a powerful and systhematical formalism to investigate quantum field theories.
Here $\varphi^i$ (resp. $\Phi^i$) denote arbitrary background (resp. bare) fields, $S[\Phi]$ is the classical action and $\mathcal{M}[\Phi]$ denotes the functional measure, to be studied in detail in Sec.~\ref{measure1}.
We also adopt the DeWitt index notation, where small letters $i=(I,x^\mu)$ account for both discrete $I$ and continuum $x^\mu$ indices. Repeated DeWitt indices thus result in summations over $I$ and integrations over spacetime.
 
At the one-loop level, Eq.~\eqref{eq:pathint} yields \cite{Barvinsky:1985an,Vassilevich:2003xt,Avramidi:2000bm}:
\begin{equation}
\Gamma[\varphi]
=
S
+ \Gamma^{(1)}
+ \Gamma^{(1)}_\mathcal{M} \, ,
\label{1loop}
\end{equation}
where
\begin{align}
	\Gamma^{(1)} &= \frac12 \Tr\log \mathcal{H}_{ij} \, ,
	\label{1loopH}
	\\
	\Gamma^{(1)}_\mathcal{M} &= - \log \mathcal{M}[\varphi] \, ,
	\label{1loopM}
\end{align}
denote the standard correction due to the Hessian
\begin{equation}
	\mathcal{H}_{ij} = \nabla_i \nabla_j S \, ,
\end{equation}
and due to the functional measure, respectively.
Note that
\begin{equation}
	\mathcal{H}_{ij}
	= \frac{\delta^2 S}{\delta\varphi^i \delta\varphi^j}
	- \Gamma^k_{ij} \frac{\delta S}{\delta\varphi^k}
	\ ,
	\label{eq:hess}
\end{equation}
where $\Gamma^k_{ij}$ denotes the configuration-space connection. Since we shall only be interested in on-shell quantities, the second term in \eqref{eq:hess} vanishes, thus the second covariant derivative reduces to the ordinary one.
The trace in Eq.~\eqref{1loopH} is to be understood in the functional sense:
\begin{equation}
	\Tr \mathcal{O}_{ij}
	=
	\int\mathrm{d}^4x
	\tr \mathcal{O}_{IJ}(x,x)
	\ ,
	\label{trace}
\end{equation}
which also includes the finite-dimensional trace $\tr$ over discrete indices. We note that the functional-measure contribution is one-loop exact.

For canonical Lagrangians, the on-shell part of the Hessian can generally be written in the minimal form: 
\begin{equation}
	\mathcal{H}_{ij}
	=
	\left[
		\delta_{IJ}(\Box-m^2)
		+ P_{IJ}
	\right]
	\delta(x,x')
	\ .
	\label{hessian}
\end{equation}
One should not confuse the $\delta_{IJ}$ in Eq.~\eqref{hessian}, which are just coefficients of some bilinear form, with the configuration-space metric $G_{IJ}$. Although the configuration-space metric is typically identified from the kinetic terms \cite{Vilkovisky:1984st,Alonso:2015fsp,Alonso:2016oah,Finn:2019aip}, this is by no means a physical requirement.
There is, in fact, no generally accepted physical reason to make such a choice \cite{BarberoG:1993cvb,Odintsov:1991yx,Huggins:1987zw,Rebhan:1986wp}. Moreover, such a procedure faces a number of difficulties when applied to fermions, which only recently has seen some progress \cite{Gattus:2023gep,Gattus:2024spj,Finn:2020nvn,Assi:2023zid}. Note also that bilinear forms are tensors in configuration space, thus they transform covariantly under field redefinitions, for
\begin{equation}
	\delta_{IJ}
	=
	\frac{\delta\widetilde\varphi^M}{\delta\varphi^I}
	\frac{\delta\widetilde\varphi^N}{\delta\varphi^J}
	h_{MN}
	\ ,
\end{equation}
in the same way a configuration-space metric would. Therefore, the kinetic term is invariant regardless the way the metric is identified:
\begin{equation}
	\mathcal L_\text{free}
	=
	\delta_{IJ}
	\partial_\mu \varphi^I \partial^\mu \varphi^J
	=
	h_{MN} \partial_\mu \widetilde\varphi^M \partial^\mu \widetilde\varphi^N
	\ ,
\end{equation}
since
\begin{equation}
	\partial_\mu \varphi^I
	=
	\frac{\partial\varphi^I}{\partial\widetilde\varphi^J}
	\partial_\mu \widetilde\varphi^J
	\ .
\end{equation}
In either case, however, this does not enforce invariance of the functional form of the action because field redefinitions are not a symmetry in general, which would require $\delta_{IJ} = h_{IJ}$.

In our case, we shall instead build the configuration-space metric using symmetry arguments (see Sec.~\ref{measure1} for more details), in very much the same way one uses effective field theory to write down Lagrangians.
From this viewpoint, along with the Lagrangian $\mathcal{L}$, the metric $G_{ij}$ must also be seen as part of the definition of the theory. The theory is thus fully specified by the ordered pair $(\mathcal L, G_{ij})$. Different metrics for the same Lagrangian would correspond to different quantum theories for the same classical one. 
Because we use effective field theory to build both $G_{ij}$ and $\mathcal L$ separately, our approach is rather conservative. It should be expected to be a good description for energies below some physical cutoff regardless of the full-fledged functional measure to be found in the UV-completed theory. In this approach, the theory's phenomenology shall depend on the free parameters appearing in $G_{ij}$, which could then be fixed by experimental data rather than by the comparison with the kinetic term.

The Hessian depends solely on the underlying gauge connections (via covariant derivatives) and interactions, thus being fully characterized by potential terms, hereby denoted $P^I_{\ J}$, and the gauge field strength (fiber bundle curvature):
\begin{equation}
	[D_\mu, D_\nu]\varphi^I = \Omega^{I}_{\ J \mu\nu} \varphi^J
    \ ,
\end{equation}
where $\Omega^{I}_{\ J \mu\nu} = -i g F^a_{\mu\nu} (T^a)^I_{\ J}$.

The core of the aforementioned approximations relies on the heat kernel $K(s) = e^{s\mathcal{H}}$, which enters the one-loop effective action through the Schwinger proper-time method \cite{Barvinsky:1985an,Vassilevich:2003xt,Avramidi:2000bm}:
\begin{equation}
\Gamma^{(1)} = -\frac12\int_{1/\Lambda}^\infty \frac{\mathrm{d}s}{s}\Tr K(s),
\label{propert}
\end{equation}
where $\Lambda$ is a UV cutoff. Because $K(s)$ solves the heat equation (with appropriate boundary conditions), the one-loop effective action can be found from its solutions. 

In general, Eq.~\eqref{propert} splits up into a divergent and a finite part:
\begin{equation}
	\Gamma^{(1)}
	=
	\Gamma^{div}
	+
	\Gamma^{fin}
	\ .
\end{equation}
Asymptotic expansions on the proper time then allows one to focus on certain regimes of interest. The standard perturbation theory regime of Feynman diagrams, for example, is obtained from the Schwinger-DeWitt expansion, which finds $K(s)$ (hence the effective action) in powers of inverse mass, namely
\begin{align}
	\frac{\mathcal{R}}{m^2} \ll 1 \ , \quad
	\frac{\nabla^2}{m^2} \ll 1 \ ,
	\label{SDpar}
\end{align}
where $\mathcal R = (P^I_{\ J}, \Omega^I_{\ J\mu\nu})$ denotes generalized curvatures~\footnote{Spacetime curvatures also show up, but we omit them since we are only interested in Minkowski backgrounds.}, which includes potentials $P^I_{\ J}$ and field strengths $\Omega^I_{\ J\mu\nu}$.
This expansion reveals both the one-loop divergence structure $\Gamma^{div}$, which is renormalized by the operators contained in the bare action $S[\varphi]$, and the physical finite predictions $\Gamma^{fin}$. Because of Eq.~\eqref{SDpar}, the Schwinger-DeWitt expansion can also be seen as a double expansion in field strengths and derivatives, thus resulting in a local finite part $\Gamma^{fin}$.

More general expansions also exist, which permit the study of quantum field theory in other regimes.
In covariant perturbation theory, for example, one gives up the second constraint of Eq.~\eqref{SDpar}. In this case,
the heat kernel can be expressed as a series in the generalized curvatures with infinitely many derivatives \cite{Barvinsky:1987uw,Barvinsky:1990up}:
\begin{align}
	\Tr K(s)
	&=
	\frac{1}{(4\pi s)^2}
	\int\mathrm{d}^4x
	\Big\{
		\delta^I_{\ I}
		+ s\,  P^I_{\ I}
		+ s^2 \left[
				P_{IJ} \, f_P(-s\Box) \, P^{IJ}
				+ \Omega_{IJ\mu\nu} \, f_\Omega(-s\Box) \, \Omega^{IJ\mu\nu}
			\right]
	\Big\}
	\label{hk}
	\\
	&+ \mathcal{O}(\mathcal{R}^3)
	\ ,
	\nonumber
\end{align}
where
\begin{align}
    f_P(\xi)
    &=
    \frac12 f(\xi)
    \ ,
    \\
    f_\Omega(\xi)
    &=
    -\frac12 \frac{f(\xi) - 1}{\xi}
    \ ,
    \\
    f(\xi)
    &=
    \int_0^1 \mathrm{d}\alpha \, e^{-\alpha(1-\alpha)\xi}
    \ .
\end{align}
From Eqs.~\eqref{propert} and \eqref{hk}, one finds~\footnote{We have dropped the zeroth and linear order terms. The former is a constant that does not affect the equations of motion and the only effect of the latter is to renormalize the global coefficient in the potential.}:
\begin{equation}
    \Gamma^{fin}
    =
    -\frac{1}{2(4\pi)^2}
    \int\mathrm{d}^4x
		\left[
				P_{IJ} \, \gamma_P\left(-\frac{\Box}{m^2}\right) \, P^{IJ}
				+ \Omega_{IJ\mu\nu} \, \gamma_\Omega\left(-\frac{\Box}{m^2}\right) \, \Omega^{IJ\mu\nu}
			\right]
	+ \mathcal{O}(\mathcal{R}^3)
	\ ,
	\label{EA1}
\end{equation}
where the form factors are given by:
\begin{align}
	\gamma_P(u)
	&=
	- \frac12 \int_0^1 \mathrm{d}\alpha 
	\log\left[1 + u \alpha (1-\alpha) \right]
	\ ,
	\\
	\gamma_\Omega(u)
	&=
	\frac{1}{12}
	- \frac12 \int_0^1 \mathrm{d}\alpha 
	\left[\frac{1}{u} + u \alpha (1-\alpha) \right]
	\log\left[1 + u \alpha (1-\alpha) \right]
	\ .
	\label{factorO}
\end{align}
One should note that Eq.~\eqref{EA1}, albeit written only in terms of background bosonic fields, also accounts for integrated-out fermions because fermions also fall in the same Hessian structure of Eq.~\eqref{hessian}.

Eq.~\eqref{EA1} is seen as a partial resummation of infinitely many terms of the form $\mathcal{R}\Box^n \mathcal{R}$ contained in the finite part of Schwinger-DeWitt method.
It is important to stress that Eq.~\eqref{EA1} takes this same form, with the same form factors, for any theory described by the minimal Hessian of Eq.~\eqref{hessian}. The differences are only reflected on their coefficients, which result from the field space where the traces of Eq.~\eqref{EA1} are taken and thus depend on the number and types of fields being integrated out in Eq.~\eqref{eq:pathint}. Regardless of the bare action one starts with, Eq.~\eqref{EA1} provides corrections to the scalar and gauge sectors. 
It is this important result that allows parameterization of new physics in terms of a handful of free parameters. Bounds on these parameters would thus restrict the number and types of fields existing in a fundamental theory.

For our purposes, let us assume the unknown bare action contains the electroweak's gauge sector:
\begin{equation}
	S
	\supset
	\int\mathrm{d}^4x
	\left[
	(D^{\mu} \varphi)^\dag(D_\mu \varphi)
	+ V(\varphi)
	+ \frac{1}{4}W_{\mu\nu}^a W^{a\mu\nu}
	+ \frac{1}{4}B_{\mu\nu}B^{\mu\nu}
	\right]
	\ ,
	\label{bareL}
\end{equation}
where
\begin{equation}
	V(\varphi)
	=
	\frac{\lambda^2}{2}(\varphi^\dag\varphi-\eta^2)^2 \, ,
\end{equation}
and $\lambda$ and $\eta$ denote the Higgs self-coupling and the vev, respectively.
The $SU(2)\times U(1)$ covariant derivative reads 
\begin{equation}
	D_\mu
	=
	\partial_\mu 
	- i g W^a_\mu T^a - i g' Y B_\mu
	\ ,
\end{equation}
where $g$ and $g'$ are, respectively, the corresponding coupling constants of the gauge fields $W^a_\mu$ and $B_\mu$ to the Higgs sector, $T^a$ is the SU(2) generator and $Y$ is the weak hypercharge.
In this case, the fiber bundle curvature accounts for both the weak and the electromagnetic field strength:
\begin{equation}
	(\Omega_{\mu\nu})^I_{\ J}
	=
	- i g W^a_{\mu\nu} (T^a)^I_{\ J}
	- i g' Y B_{\mu\nu} \delta^I_{\ J}
	\ .
\end{equation}
For each field in the fundamental Lagrangian, there would be a correction of the type \eqref{EA1}. All these possible contributions add up to wit~\footnote{We note that cross-terms in Eq.~\eqref{LEW} vanish identically because they are proportional to the trace of the $SU(2)$ generators, which is always zero.}:
\begin{align}
	\mathcal{L}^{(1)}_{EW}
    =
    \sum_{i=1}^N
    \Bigg[
    	&
		c_{\varphi} V''(\varphi) \gamma_P\left(-\frac{\Box}{m_{\varphi_i}^2}\right) V''(\varphi)
		+ c_W W^a_{\mu\nu} \gamma_\Omega\left(-\frac{\Box}{m_i^2}\right) W^{a\mu\nu}
		\label{LEW}
		\\
		&
		+ c_{B_i} B_{\mu\nu} \gamma_\Omega\left(-\frac{\Box}{m_i^2}\right) B^{\mu\nu}
	\Bigg]
	\ ,
	\nonumber
\end{align}
where $m_{\varphi_i}, m_i$ are the masses of the particles, present in the fundamental Lagrangian, running in the loop and
\begin{align}
	c_\varphi
	&=
	-\frac{1}{2(4\pi)^2} \, ,
	\\
	c_W
	&=
	\frac{g^2}{4(4\pi)^2} \, ,
	\\
	c_{B_i}
	&=
	\frac{g'^2}{2(4\pi)^2} Y_i^2 n_i
	\ .
\end{align}
Here $N$ denotes the number of particles in the fundamental theory, $Y_i$ and $n_i$ are the weak hypercharge and the dimension of the representation, respectively, for each of these particles. The effects of BSM fields and interactions present in the bare action are fully accounted by $c_{\varphi},c_W,c_{B_i}, m_{\varphi_i}$ and $m_i$. Moreover, only $c_{B_i}$ carries information on the fundamental particle's spin, $c_\varphi$ and $c_W$ are the same for any spin.

\section{The functional measure in the electroweak sector}
\label{measure1}

The functional measure $\mathcal M[\Phi]$ in Eq.~\eqref{eq:pathint} plays a fundamental role in physics. Examples of its uses include the study of anomalies \cite{Fujikawa:1979ay,Fujikawa:1980eg,Fujikawa:1980vr}, the cancelation of ultralocal divergences \cite{Fradkin:1973wke,Fradkin:1976xa} and the formal justification for the Fadeev-Poppov method \cite{Babelon:1979wd,Babelon:1980uj,Ellicott:1990up,Ellicott:1989mi}. The functional measure has also been studied in curved spacetimes, where its presence is important for a correct descripition of quantum gravity \cite{DeWitt:1967ub,Fujikawa:1983im,Bern:1990bh,Anselmi:1991wb,Ohta:2016npm,Ohta:2016jvw}.

Although some dispute exists regarding its form \cite{Unz:1985wq,Toms:1986sh,Moretti:1997qn,Hatsuda:1989qy,vanNieuwenhuizen:1989dx,Armendariz-Picon:2014xda,Becker:2020mjl,Buchbinder:1987vp,Hamamoto:2000ab}, from a geometrical perspective $\varphi^i$ are mere configuration-space coordinates, thus a natural definition for the measure would be \cite{Mottola:1995sj,DeWitt:2003pm,Toms:1986sh,Casadio:2022ozp}:
\begin{equation}
    \mathcal M[\varphi] = \sqrt{\Det G_{ij}}
    \ ,
    \label{measure}
\end{equation}
where $\Det G_{ij}$ denotes the functional determinant of the configuration-space metric $G_{ij}$. This definition makes sense as the functional generalization of integrations on manifolds. It is also appealing, for it upholds, together with a connection compatible to $G_{ij}$, invariance under field redefinitions and gauge transformations \cite{DeWitt:2003pm,Vilkovisky:1984st}. Indeed, Eq.~\eqref{measure} transforms a (functional) scalar density which cancels the Jacobian from $\mathcal{D}\varphi$. The phenomenological consequences of the functional measure \eqref{measure} has just started to be investigated \cite{Casadio:2022ozp,Kuntz:2022kcw,deFreitas:2023ujo,Casadio:2024vfh}.

As it has already been stressed, there is no canonical choice for $G_{ij}$, which must be seen as part of the definition of the theory. Different configuration-space metrics correpond to unequivalent quantization schemes. Thus to fully specify the quantum theory, one must provide the classical action furnished with some $G_{ij}$. One popular approach defines $G_{ij}$ from the Lagrangian's kinetic terms \cite{Vilkovisky:1984st,Alonso:2015fsp,Alonso:2016oah,Finn:2019aip}. However, this is an ad-hoc procedure built upon aesthetical (rather than physical) motivations \cite{BarberoG:1993cvb,Odintsov:1991yx,Huggins:1987zw,Rebhan:1986wp}.
Indeed, the formal equivalence of the path integral and canonical quantization holds for any choice of $G_{ij}$ should the phase-space measure be correctly identified (see the Appendix \ref{app}).

Another approach involves imposing ultralocality, i.e., proportionality with the Dirac delta (but not its derivatives):
\begin{equation}
	G_{ij}
	=
	G_{IJ}(\varphi)
	\,
	\delta^{(4)}(x,x') \, ,
	\label{ultralocal}
\end{equation}
and requiring, for simplicity, that no dimensionful parameters are introduced in $G_{IJ}$ \cite{Toms:2009vd,Parker:2009uva,Percacci:2017fkn,DeWitt:2003pm,Casadio:2021rwj}.
Notice that while $G_{ij}$ is a metric functional in an infinite-dimensional space, $G_{IJ}(\varphi)$ is a metric function (not functional) of the fields (but not their derivatives) in a finite-dimensional space. The Dirac delta in Eq.~\eqref{ultralocal} enforces the same $G_{IJ}$ across all spacetime points and prevent non-local effects.

For all interactions but gravity, both approaches provide exactly the same metric with no free parameters.
In quantum gravity, on the other hand, the second approach yields a one-parameter family of metrics, collectively called DeWitt metric, which depends on a free dimensionless parameter \cite{DeWitt:1967yk}. The dependence on this parameter could be fixed from the kinetic term (as per the first approach), in which case different theories would provide different values for it, but otherwise it can be left arbitrary and fixed by physical requirements \textit{a posteriori}. In Ref.~\cite{Odintsov:1991yx}, for example, the dependence of the radii of spontaneous compactification on the DeWitt parameter has been studied. In Ref.~\cite{BarberoG:1993cvb}, the dependence of the renormalization group equations on the DeWitt parameter is used to obtain sensible physical results, which leads to interesting phenomenology regarding the cosmological constant decay and dark matter. In Ref.~\cite{Huggins:1987zw}, a well-defined effective action in spacetimes with compact higher dimensions led to stringent bounds on the DeWitt parameter. More recently, the dependence on the DeWitt parameter has also been investigated in relation to the conformal factor problem in gravitational path integrals \cite{Liu:2023jvm}.

Imposing the absence of dimensionful parameters avoids higher-order field-dependent contributions in $G_{IJ}$, which are not actually prohibitted by any physical principle.
In fact, one could only require the invariance of $G_{IJ}(\varphi)$ under the same symmetries present in the Lagrangian and write it as an expansion in inverse powers of the cutoff, following the effective field theory realm.
At leading order, this amounts on the aforementioned natural choices that do not introduce dimensionful parameters, as adopted in Refs.~\cite{Toms:2009vd,Parker:2009uva,Percacci:2017fkn,DeWitt:2003pm,Casadio:2021rwj}. Apart from the gravitational case, which leads to the DeWitt metric, the natural choices (no dimensional parameters) for all other interactions result in field-independent, and hence flat, metrics.
Beyond leading order, however, new dimensionful~\footnote{Dimensionful parameters can always be traded by dimensionless ones by dividing the latter by an appropriate power of the cutoff $\Lambda$.} free parameters show up as coefficients of higher-order operators (see Eq.~\eqref{detG} below). These higher-order operators contribute to the configuration-space curvature, albeit as only small deviations from the flat one due to the small corrections in effective field theory (large cutoff).
The new parameters cannot be computed from first principles within the effective field theory and are supposed to be fixed by experiments. We should stress that, while the definition of the configuration-space metric as independent of the Lagrangian is well-known \cite{BarberoG:1993cvb,Odintsov:1991yx,Huggins:1987zw,Rebhan:1986wp}, this is the first time that the effective field theory is recognized as a tool to explore the configuration-space geometry.

As pointed out before, we also stress that, regardless of whether the metric is identified from the kinetic term, the functional form of the action is not invariant under field redefinitions. The Vilkovisky--DeWitt effective action takes different forms in different configuration-space charts, but it is invariant in the sense of a scalar functional~\footnote{The classical action is automatically invariant in this sense without the need of introducing additional structures.} \cite{DeWitt:2003pm,Vilkovisky:1984st}:
\begin{equation}
	\widetilde \Gamma[\widetilde\varphi]
	=
	\Gamma[\varphi]
	\ ,
\end{equation}
where $\widetilde\Gamma[\widetilde\varphi]$ denotes the transformed action evaluated at the transformed field $\widetilde\varphi^i = \widetilde\varphi^i(\varphi)$.
For this reason, field redefinitions are not a symmetry of the action in the usual sense, hence $\Gamma[\widetilde\varphi]	\neq	\Gamma[\varphi]$. Solutions are taken into solutions because the functional derivative of the action transforms covariantly:
\begin{equation}
	\frac{\delta\widetilde\Gamma[\widetilde\varphi]}{\delta\widetilde\varphi^i}
	=
	\frac{\delta\varphi^j}{\delta\widetilde\varphi^i}\frac{\delta\Gamma[\varphi]}{\delta\varphi^j}
	\ ,
\end{equation}
for invertible transformations.

From Eqs.~\eqref{1loopM}, \eqref{trace} and \eqref{ultralocal}, one can see that ultralocality is responsible for extreme divergences $\delta^{(4)}(0)$ in the quantum action. The effective field theory lore is to declare our ignorance of the UV by imposing a physical cutoff $\Lambda$, which makes the theory finite and sets the scale of new physics. We can implement such a cutoff via a Gaussian regularization:
\begin{equation}
	\delta^{(4)}(x)
	=
	\frac{\Lambda^4}{(2\pi)^{2}} e^{\frac{-x^2 \Lambda^2}{2}}
	\ ,
	\label{ultradiv}
\end{equation}
hence $\delta^{(4)}(0) = \frac{\Lambda^4}{(2\pi)^{2}}$.
From Eqs.~\eqref{1loopM} and \eqref{measure}--\eqref{ultradiv}, we then find:
\begin{equation}
    \Gamma^{(1)}_\mathcal{M} 
    =
    - \frac{\Lambda^4}{8\pi^{2}}
    \int\mathrm{d}^4x
    \log\det G_{IJ}
	\ ,
	\label{genact}
\end{equation}
Note that the Dirac delta divergence is polynomial. The correction \eqref{genact} is thus expected to be UV sensitive.

For the electroweak sector, there is no gauge-invariant combination of the gauge fields without derivatives, hence $G_{IJ}$ must depend only on the Higgs field. At second order, the most general metric determinant reads:
\begin{equation}
	\det G_{IJ}
	=
	A + B \frac{\varphi^\dag \varphi}{\Lambda^2}
	+ \mathcal{O}(\varphi^3)
	\ ,
	\label{detG}
\end{equation}
where $A$ and $B$ are dimensionless free parameters. Eq.~\eqref{genact} then becomes:
\begin{equation}
	\Gamma^{(1)}_\mathcal{M} 
    =
    - \frac{\Lambda^4}{8\pi^{2}}
    \int\mathrm{d}^4x
    \log(A + B \frac{\varphi^\dag \varphi}{\Lambda^2})
    \ ,
    \label{looppot}
\end{equation}
which provides another correction to the Higgs effective potential.

Finally, we point out the differences and similarities between our approach to the existing ones in the literature. Geometrical tools have also been devised to study the electroweak sector in the context of the Higgs effective field theory \cite{Alonso:2015fsp,Alonso:2016oah,Alonso:2023upf,Alonso:2021rac,Alonso:2016btr,Alonso:2022ffe,Cohen:2020xca,Cohen:2021ucp,Cohen:2022uuw,Helset:2022pde,Helset:2022tlf}. In both approaches, the only assumption about the fundamental spectra regards the presence of the standard model fields, namely the electroweak gauge fields and the Higgs (see Eq.~\eqref{bareL}). However, kinetic terms (and interactions involving derivatives) may differ across these approaches. Indeed, because in the Higgs effective field theory the configuration-space metric is extracted from the Lagrangian, one cannot obtain the potential correction \eqref{looppot} without simultaneously modifying the kinetic term in the Higgs effective field theory.
Changing kinetic terms also change the propagators and their pole structure, which brings important differences with respect to our approach, such as the value of the physical masses. 
On the other hand, the functional measure in our case affects only the potential terms. 
Disentangling the metric from the Lagrangian allows for a more general approach not covered by the Higgs effective field theory.
This also facilitates the generalization to non-scalar sectors without facing issues with tachyon or ghost instabilities, which would be rather non-trivial otherwise.

\section{Higgs mechanism and oblique corrections}
\label{Higgs}

The effective action above, including the definition of the functional measure, were obtained in the Euclidean space. To obtain physical results, we must return to Minkowski signature. From now on, everything shall be written in Minkowski space with signature $(+---)$.

The first term in Eq.~\eqref{LEW} does not contribute to the Higgs potential since it contain derivatives.
The total one-loop Higgs effective potential thus receives corrections only from the functional measure~\footnote{Corrections of the Coleman-Weinberg type takes place when one gives up the first (rather than the second) condition in Eq.~\eqref{SDpar}, which results in local terms with infinitely many generalized curvatures. We shall here consider only covariant perturbation theory where such corrections do not occur.}:
\begin{equation}
	\label{effpot}
	V_{eff}
	=
	\frac{\lambda^2}{2} (\varphi^\dag\varphi-\eta^2)^2
	- \frac{\Lambda^4}{8\pi^{2}} \log(1 + C \frac{\varphi^\dag \varphi}{\Lambda^2})
	\ ,
\end{equation}
where we defined the new constant $C=B/A$. This potential determines the effective Higgs mass:
\begin{equation}
	M_H
	=
	\sqrt{
		3 \lambda^2 \varphi_0^2
		- \lambda^2 \eta^2
		- \frac{\Lambda^4 C}{8\pi^{2} (1+C\varphi_0^2)}
		+ \frac{\Lambda^4 C^2\varphi_0^2}{4 \pi^{2} (1+C\varphi_0^2)^2}
	}
	\ .
\end{equation}

The masses of the $W$ and $Z$ bosons, on the other hand, are obtained from the Higgs mechanism, as usual.
The minimum of Eq.~\eqref{effpot} yields the corrected vacuum expectation value~\footnote{There are two roots when solving for the minimum. However, only the result in \eqref{eq619} returns to the classical value in the limit $\hbar\to 0$.}:
\begin{equation}
	\label{eq619}
	\varphi_0^2
	\equiv
	\varphi^\dag \varphi
	= 
	\eta^2
	+ \frac{\Lambda^2 C}{8\pi^{2} \lambda^2 \left(1+\frac{C\eta^2}{\Lambda^2}\right)}
	+ \mathcal{O}(\hbar^2)
	\ .	
\end{equation}
We note that the functional measure shifts the Higgs vacuum expectation value $\eta\to\varphi_0$.
 
After symmetry breaking, Eqs.~\eqref{bareL} and \eqref{LEW} yield the effective Lagrangian for the electroweak interaction:
\begin{align}
	\mathcal{L}_{1PI}
	&=
	W^+_{\mu} \, 
	\left[
		\Pi^{\mu\nu}_W(\Box) 
		+ M_{W,0}^2 \eta^{\mu\nu} 
	\right] W^{-}_{\nu}
	+
	\frac12 W^3_{\mu} \, 
	\left[
		\Pi^{\mu\nu}_W(\Box) 
		+ M_{Z,0}^2 \cos^2\theta_W \eta^{\mu\nu}
	\right] W^3_{\nu}
	\\
	&+ \frac12 B_{\mu} \, 
	\left[
		\Pi^{\mu\nu}_B(\Box)
		+ M_{Z,0}^2 \sin^2\theta_W \eta^{\mu\nu}
	\right] B_{\nu}
	\nonumber
	- M_{Z,0}^2 \cos\theta_W \sin\theta_W \, W_\mu^3  B^\mu
	\ ,
\end{align}
where we made the usual definitions:
\begin{align}
	&\bar g = \sqrt{g^2 + g'^2} \ ,
	\\
	&\cos\theta_W = \frac{g}{\bar g} \ ,
	\\
	&\sin\theta_W = \frac{g'}{\bar g} \ ,
	\\
	&M_{W,0}
	=
	\frac{g}{\sqrt{2}}\varphi_0
	\label{MW}
	\ ,
	\\
	&M_{Z,0}
	=
	\frac{\bar g}{\sqrt{2}}\varphi_0
	\ ,
	\label{MZ}
\end{align}
and~\footnote{Note the different sign inside the form factors in Minkowski space in contrast to the Euclidean form factors of Sec.~\ref{EW}.}
\begin{equation}
	\Pi^{\mu\nu}_K(\Box)
	=
		\eta^{\mu\nu}
		\left[
			\left(1 + 4 \sum_i c_{K_i} \gamma_\Omega\left(\frac{\Box}{m_{i}^2}\right) \right) \Box
		\right]
		-
		\left[
			1
			+ 4 \sum_i c_{K_i} \gamma_\Omega\left(\frac{\Box}{m_{i}^2}\right)
		\right]
		\partial^\mu \partial^\nu
		\ ,
		\label{totalform}
\end{equation}
for $K = W,B$. The Lagrangian $\mathcal{L}_{1PI}$ corresponds to the gauge sector of the effective action $\Gamma$, whereas the Higgs sector has been omitted for simplicity.

To identify the boson masses, one must diagonalize the quadratic terms.
Because of the different coefficients $c_W \neq c_{B_i}$ of the non-local terms in Eq.~\eqref{totalform}, the corresponding form factors of each interaction differ from each other. As a result, a rotation by $\theta_W$ can no longer diagonalize the quadratic terms. Such diagonalization requires a non-local linear transformation, which at one-loop level reads~\footnote{This can be easily obtained by performing a general linear transformation and setting the coefficients of the cross-terms so obtained to zero. We also imposed that $A_\mu$ be massless and the $Z$ boson mass be given by its usual expression \eqref{MZ}.}:
\begin{align}
	\renewcommand*{\arraystretch}{1.5}
	\begin{pmatrix}
		W^3_\mu
		\\
		B_\mu
	\end{pmatrix}
	=
	\begin{pmatrix}
		\left(1 - g(\Box) \sin^2\theta_W \right) \cos\theta_W & \sin\theta_W \\
		- \left( 1 + g(\Box) \cos^2\theta_W \right) \sin\theta_W & \cos\theta_W
	\end{pmatrix}
	\begin{pmatrix}
		Z_\mu \\
		A_\mu
	\end{pmatrix} \, ,
	\label{diag}
\end{align}
where
\begin{equation}
	g(\Box)
	=
	4 \sum_i (c_W - c_{B_i}) \, \gamma_\Omega\left(\frac{\Box}{m_{i}^2}\right)
	\ .
\end{equation}
Note that $g(\Box) \to 1$ for $c_{B_i} = c_W$ or at the classical limit $\hbar\to 0$ since the form factor is of one-loop order, thus recovering Weinberg's rotation by $\theta_W$.

Under the transformation \eqref{diag}, the Lagrangian transforms to:
\begin{equation}
	\mathcal{L}_{1PI}
	=
	W^+_{\mu} \left[
		\Pi^{\mu\nu}_W(\Box) 
		+ M_{W,0}^2 \eta^{\mu\nu}
	\right] W^{-}_{\nu}
	+
	\frac12 Z_{\mu} 
	\left[
		\Pi^{\mu\nu}_Z(\Box)
		+ M_{Z,0}^2 \eta^{\mu\nu}
	\right] Z_{\nu}
	+ \frac12 A_{\mu} \Pi^{\mu\nu}_A(\Box) A_{\nu}
	\label{effEW}
	\ ,
\end{equation}
with
\begin{align}
	\Pi^{\mu\nu}_Z(\Box)
	&=
	\cos^2\theta_W \left(1 - 2 \sin^2 \theta_W g(\Box) \right) \Pi^{\mu\nu}_W(\Box)
	+ \sin^2\theta_W \left(1 + 2 \cos^2 \theta_W g(\Box) \right) \Pi^{\mu\nu}_B(\Box)
	\ ,
	\\
	\Pi^{\mu\nu}_A(\Box)
	&=
	\sin^2\theta_W \Pi^{\mu\nu}_W (\Box)
	+ \cos^2\theta_W \Pi^{\mu\nu}_B (\Box)
	\ .
\end{align}

From Eq.~\eqref{effEW}, we can identify the mass poles in momentum space:
\begin{align}
	\label{loop47}
	&\left[
		1
		+ 4 \, c_W \sum_i \gamma_\Omega \left( \frac{-k^2}{m_{i}^2} \right)
	\right] k^2
	- M^2_{W,0}
	= 
	0
	\ ,
	\\
	&\left[
		1
		+ 4 \sum_i \gamma_\Omega \left( \frac{-k^2}{m_{i}^2} \right) \left(c_W \cos^2\theta_W + c_{B_i} \sin^2\theta_W\right)
	\right] k^2
	- M^2_{Z,0}
	=
	0
	\ ,
\end{align}
from which
one can read off the physical masses $M_{W}$ and $M_Z$, defined as the poles of the corresponding propagators at $k=M_{W}, M_Z$:
\begin{align}
	\label{loop48}
	M_W^2
	&=
	M_{W,0}^2
	- 4 \, c_W M_W^2 \sum_i \gamma_\Omega \left( 
		\frac{-M_W^2}{m_{i}^2} 
	\right) \, ,
	\\
	M_Z^2
	&=
	M_{Z,0}^2
	- 4 \, M_Z^2 \sum_i \gamma_\Omega \left( \frac{-M_Z^2}{m_{i}^2} \right) \left(c_W \cos^2\theta_W + c_{B_i} \sin^2\theta_W\right)
	\ .
\end{align}
Since the form factor $\gamma_\Omega(u)$ is already of one-loop order, to this order we can take $M_{i} \approx M_{i,tree}$ on the RHS, hence:
\begin{align}
	\label{MW1loop}
	M_W^2
	&=
	M_{W,tree}^2
	\left[
		1
		- 4 \, c_W \sum_i \gamma_\Omega \left( 
		\frac{-M_{W,tree}^2}{m_{i}^2} 
	\right)
	+ \frac{\Lambda^2 C}{8\pi^2 \lambda^2 \eta^2}
	\right]
	\ ,
	\\
	\label{MZ1loop}
	M_Z^2
	&=
	M_{Z,tree}^2
	\left[
		1
		- 4 \sum_i \gamma_\Omega \left( \frac{-M_{Z,tree}^2}{m_{i}^2} \right) \left(c_W \cos^2\theta_W + c_{B_i} \sin^2\theta_W\right)
		+ \frac{\Lambda^2 C}{8\pi^2 \lambda^2 \eta^2}
	\right]
	\ ,
\end{align}
where the tree-level masses read:
\begin{align}
	M_{W,tree}
	&=
	\frac{g}{\sqrt{2}}\eta
	\ ,
	\\
	M_{Z,tree}
	&=
	\frac{\bar g}{\sqrt{2}}\eta
	\ .
\end{align}
We note that the usual relation between the boson masses and the Weinberg angle $\theta_W$ no longer holds in general:
\begin{equation}
	\rho \equiv \frac{M_W^2}{M_Z^2 \cos^2\theta_W} \neq 1
	\ ,
\end{equation}
where the $\rho$-parameter is given by:
\begin{align}
	\rho
	=
	\frac{
		1
		- 4 \, c_W \sum_i \gamma_\Omega \left( 
		-\frac{M_{W,tree}^2}{m_{i}^2} 
		\right)
		+ \frac{\Lambda^2 C}{8\pi^2 \lambda^2 \eta^2}
	}
	{
		1
		- 4 \sum_i \gamma_\Omega \left( \frac{-M_{Z,tree}^2}{m_{i}^2} \right) \left(c_W \cos^2\theta_W + c_{B_i} \sin^2\theta_W\right)
		+ \frac{\Lambda^2 C}{8\pi^2 \lambda^2 \eta^2}
	}
	\ .
	\label{rho}
\end{align}
This alleviates the standard model tight constraint and allows for the accommodation of possible experimental anomalies reflecting BSM physics.

The form factor in Eq.~\eqref{rho}, given by the integral form in Eq.~\eqref{factorO}, contains all one-loop effects at second order in the field strength with infinitely many derivatives. Such generality makes it difficult to obtain useful bounds from the experimental value for $\rho$, namely \cite{Workman:2022ynf}:
\begin{equation}
	\label{rhoexp}
	\rho = 1.00038
	\ .
\end{equation}
Further assumptions about the fundamental degrees of freedom, particularly regarding their masses, allow us to massage the form factor into simpler expressions. We shall now analyse a few different regimes.

\subsection{Heavy fields}

For $m_i \gg M_{i,tree}$, the form factor \eqref{factorO} can be expanded as \cite{Barvinsky:1987uw,Barvinsky:1990up,Codello:2015mba}:
\begin{equation}
	\gamma_\Omega(u)
	=
	-\frac{u}{120}
	+ \frac{u^2}{1680}
	+ \mathcal{O}(u^3)
	\ .
\end{equation}
This is the usual regime of effective field theory, where heavy fields decouple, affecting the IR only through local operators \cite{Appelquist:1974tg}. In this case, to leading order, the masses \eqref{MW1loop}-\eqref{MZ1loop} read:
\begin{align}
	M_W^2
	&=
	M_{W,tree}^2
	\left[
		1
		+ \frac{c_W}{30} \sum_i
			\left( 
				\frac{M_{W,tree}}{m_{i}} 
			\right)^2
	+ \frac{\Lambda^2 C}{8\pi^2 \lambda^2 \eta^2}
	\right]
	\ ,
	\\
	M_Z^2
	&=
	M_{Z,tree}^2
	\left[
		1
		+ \frac{1}{30} \sum_i \left(c_W \cos^2\theta_W + c_{B_i} \sin^2\theta_W\right)
		\left(
			\frac{M_{Z,tree}}{m_{i}}
		\right)^2
		+ \frac{\Lambda^2 C}{8\pi^2 \lambda^2 \eta^2}
	\right]
	\ ,
\end{align}
and Eq.~\eqref{rho} becomes:
\begin{equation}
	\rho
	=
	1
	-
	\frac{
		\sin^2\theta_W M_{Z,tree}^2
	}{
		30\left(1 + \frac{\Lambda^2 C}{8\pi^2 \lambda^2 \eta^2}\right) 
	}
	\sum_i
	\frac{c_{B_i}}{m_{i}^2}
	+ \mathcal{O}\left(m_i^{-4}\right)
	\ .
\end{equation}
We note that $c_W$ drops out, thus the above expression sets a bound on the combination $c_{B_i}/m_i$ only.
The effects of the functional measure are contained in the dimensionless parameter $C$. The limit $C\to 0$ corresponds to a flat configuration space, in which case the functional measure is trivial. In this case, the only effect is due to the one-loop form factor from the Barvinsky-Vilkovisky resummation.  
From Eq.~\eqref{rhoexp} and the experimental values for the other parameters \cite{Workman:2022ynf}:
\begin{align}
	\label{gexp}
	g &= 0.63 ,
	\\
	\bar g &= 0.72 ,
	\\
	\eta &= 173.95 \, \text{GeV},
	\\
	\label{lambdaexp}
	\lambda &= 0.51,
\end{align}
we find
\begin{equation}
	\sum_i \frac{c_{B_i}}{m_{i}^2}
	= - 6.32 \times 10^{-6} \, \text{GeV}^{-2}
	\ .
	\label{ct1}
\end{equation}
On the other hand, because $C$ is dimensionless, one would actually expect $C\sim 1$, hence the strength of the functional measure contribution is completely controlled by the physical cutoff $\Lambda$. 
Assuming $C$ of order one and $\Lambda \sim 1$ TeV, we rather obtain:
\begin{equation}
	\frac{1}{C} \sum_i \frac{c_{B_i}}{m_{i}^2}
	= 
	- 1.65 \times 10^{-5}
	\,
	\text{GeV}^{-2}
	\ .
	\label{ct2}
\end{equation}
Eqs.~\eqref{ct1} and \eqref{ct2} offer important constraints on UV physics. Should an independent measurement be performed to determine the coefficients $c_{B_i}$, one could predict the masses of a fundamental model.

\subsection{Light fields}
In the opposite limit, where the fundamental fields are nearly massless $m_i \ll M_{i,tree}$, the form factor becomes \cite{Barvinsky:1987uw,Barvinsky:1990up,Codello:2015mba}:
\begin{equation}
	\gamma_\Omega(u)
	=
	\frac{2}{9}
	- \frac{1}{12} \log(u)
	+ \mathcal{O}(u^{-1})
	\ .
	\label{masslesslimit}
\end{equation}
In particular, in the strict massless limit, all higher order terms in \eqref{masslesslimit} vanish identically.
In this case, and assuming the size of the form factor becomes larger than the functional measure contribution, 
Eqs.~\eqref{MW1loop}, \eqref{MZ1loop} and \eqref{rho} take the form~\footnote{We recall that only the real part of the mass is interpreted as the physical mass, whereas its imaginary part (due to the negative argument in the logarithmic form factor) measures the particle decay. We have thus taken the absolute value of the argument of the logarithm.}:
\begin{align}
	M_W^2
	&=
	M_{W,tree}^2
	\left[
		1
		- \frac{8}{9} N c_W  
		+ \frac{N c_W}{3}
		\log \left( 
		\frac{M_{W,tree}^2}{\mu^2} 
	\right)
	\right]
	\ ,
	\\
	M_Z^2
	&= 
	M_{Z,tree}^2
	\left[
		1
		- \left(
			\frac{8}{9}
			- \frac{1}{3} \log \left( 
				\frac{M_{Z,tree}^2}{\mu^2} 
			\right)
		\right) \sum_i (c_W \cos^2\theta_W + c_{B_i} \sin^2\theta_W)
	\right] \, ,
	\\
	\rho
	&=
	\label{rhomassless}
	\frac{1}{9}
	- \frac{8}{9} N c_W
	- \frac{1}{3} \left[\sum_i c_{B_i} \frac{\sin^4\theta_W}{\cos^2\theta_W} - N c_W \cos^2\theta_W\right]
	\log \left( 
				\frac{M_{W,tree}^2}{\mu^2} 
	\right)
	+ \mathcal{O}\left((g'/g)^4\right)
	\ ,
\end{align}
where in the last equation we used
\begin{equation}
	\log \left(
		\frac{M_{Z,tree}^2}{\mu^2}
	\right)
	=
	\log \left(
		\frac{M_{W,tree}^2}{\mu^2}
	\right)
	\frac{g'^2}{g^2}
	+ \mathcal{O}\left((g'/g)^4\right)
	\ ,
\end{equation}
which is justifiable since $g'/g \sim 10^{-1}$. Here $\mu$ is an artificial scale used to make the logarithm dimensionless.
In the strict massless limit, the divergence structure correlates with the logarithmic form factor in such a way that the running of the couplings $c_{W}(\mu)$ and $c_{B_{i}}(\mu)$ cancels out the $\mu$ dependence in the logarithm. This is the typical situation of quantum field theory, in which large logarithms are controlled by the renormalization group. On the other hand, for massive particles, however light they are, higher orders in Eq.~\eqref{masslesslimit} also contribute and no such correlation exists and the situation is more difficult.

In the strict massless limit, the experimental values~\eqref{rhoexp} and \eqref{gexp}--\eqref{lambdaexp} give:
\begin{equation}
	\sum_i c_{B_i}(M_{Z,tree}) 
	=
	1.46 \times10^{2}
	+ 9.85 \times 10^{-2} N
	\ ,
\end{equation}
where we evaluated the coefficients at the $Z$ boson mass because that is the scale at which the experimental values are obtained.
By measuring $c_{B_i}$, one is thus able to determine the number of fields $N$ in the fundamental theory.
In particular, assuming the coefficients $c_{B_i}$ to be all of the same order $c_{B_i} \sim c_{B}$ results in:
\begin{equation}
	c_{B}(M_{Z,tree}) 
	=
	\frac{1.46 \times 10^{2}}{N}
	+ 9.85 \times10^{-2}
	\ .
\end{equation}
In the large $N$ approximation, this leads to $c_{B}(M_{Z,tree}) = 9.85 \times10^{-2}$.

\subsection{Electroweak masses}
The borderline between the regimes outlined above lies at $m_i \sim M_{i,tree}$, in which case (see Eq.~\eqref{factorO}):
\begin{equation}
	\gamma_\Omega(1)
	=
	- 7.8 \times 10^{-3} \, ,
\end{equation}
and
\begin{align}
	M_W^2
	&=
	M_{W,tree}^2
	\left[
		1
		+ 3.12 \times 10^{-2} c_W N 
		+ \frac{\Lambda^2 C}{8\pi^2 \lambda^2 \eta^2}
	\right]
	\ ,
	\\
	M_Z^2
	&=
	M_{Z,tree}^2
	\left[
		1
		+ 3.12 \times 10^{-2} \left(c_W N \cos^2\theta_W + \sin^2\theta_W \sum_i c_{B_i} \right)
		+ \frac{\Lambda^2 C}{8\pi^2 \lambda^2 \eta^2}
	\right]
	\ .
\end{align}
In this case, we find:
\begin{equation}
	\rho 
	= 
	1
	+ 3.12 \times 10^{-2} \sin^2\theta_W \left(c_W N - \sum_i c_{B_i} \right)
	\ ,
\end{equation}
which, under Eq.~\eqref{rhoexp}, translates into
\begin{equation}
	\sum_i c_{B_i} = -0.05 + 6.28 \times 10^{-4} N
	\ .
\end{equation}

\subsection{Large $\Lambda$ expansion}
The limit $\Lambda\to\infty$ of Eq.~\eqref{rho} results in $\rho = 1$ as it should. Indeed, since $\Lambda$ is the scale where new physics kicks in, sending it to infinity is the same as not having new physics beyond the standard model, hence one recovers $\rho=1$. To avoid fine-tuning issues with the Higgs mass, one usually expects $\Lambda\sim 1$ TeV, which would already provide a good (large) expansion parameter:
\begin{equation}
	\frac{\Lambda^2}{\eta^2} \sim 10^2
	\ .
\end{equation}
The confirmation of the $W$ boson mass obtained at the Fermilab would indeed set the cutoff at the TeV scale. On the other hand, the LHC has not found any sign of new physics at the TeV region so far, which might suggest that the cutoff is even higher $\Lambda \gtrsim 1$ TeV and the expansion even better.  
Indeed, there is also the possibility that new physics only sets in at the GUT scale $\Lambda \sim 10^{16}$, where the renormalization group of all interactions meet.

Expanding Eq.~\eqref{rho} around $\Lambda=\infty$ gives:
\begin{align}
	\rho
	&=
	1
	+ \frac{32 \pi^2 \lambda^2 \eta^2}{C \Lambda^2}
	\sum_i
	\left[
		c_W \gamma_\Omega\left(\frac{-M_{Z,tree}^2}{m_i^2}\right)
		- \left(c_W \cos^2\theta_W + c_{B_i} \sin^2\theta_W\right) \gamma_\Omega\left(\frac{-M_{W,tree}^2}{m_i^2}\right)
	\right]
	\label{largeL}
	\\
	&+ \mathcal{O}(\Lambda^{-4})
	\ .
	\nonumber
\end{align}
The form factor is still quite difficult to deal with in general, but one can combine Eq.~\eqref{largeL} with the other aforementioned expansions to obtain more transparent expressions. In any case, Eq.~\eqref{largeL} parameterizes new physics as an inverse power series in $\Lambda$ in terms of calculable form factors.
This should not be confused with the usual procedure for writing down an effective Lagrangian in terms of local operators. Indeed, the result \eqref{largeL} provides a parameterization of non-local form factors.
We also note that $\Lambda$ only shows up due to the correct treatment of the functional measure, thus such parameterization is only possible when the functional measure is present.

\section{Conclusions}
\label{conc}
Despite the many phenomena that the standard model cannot account for, the LHC data frustatingly confirms its predictions. The CDF collaboration, on the other hand, has found evidence for new physics due to the 7$\sigma$ deviation observed in the $W$ mass, thus contradicting the LHC findings. At this stage, effective field theory seems to be the best tool for theoretically probing new physics while we await confirmation from precision tests.

In this paper, we adopted the effective action formalism as a steppingstone to effective field theory in order to parameterize the UV physics. This is based on the observation that the Barvinsky--Vilkovisky resummation provides the same form factors for any canonical Lagrangian whose Hessian involves only minimal operators. Our ignorance of the UV is then parameterized by the unknown coefficients of the non-local form factors. These reflect the unknown number and type of fields in the fundamental bare action, as described by the weak hypercharge and the dimension of the representation to which the field belongs.

Covariant perturbation theory does not give any contribution to the Higgs effective potential, but the functional measure does. From a geometrical viewpoint, the latter is written in terms of the configuration-space metric, which must be determined by symmetry arguments. We computed the Higgs effective potential with the functional measure correction at leading order.  Ultimately, this correction changed the Higgs vev and, after symmetry breaking, shifted the classical values of the $W$ and $Z$ boson masses. The joint contributions from the functional measure and the non-local form factor to the inverse dressed propagator have been obtained, from which the most general masses, at one-loop, have been read off. 

Our result alleviates the tight relation between $M_Z$, $M_W$ and $\rho$ found at the standard model, thus allowing for the description of the deviation discovered by Fermilab, should it be confirmed. On the other hand, the comparison with experimental data has provided important constraints on the UV spectra. In all cases, there exist constraints relating the number of fields, the scale of new physics and the details of the fundamental particles (e.g. masses, hypercharge). 
In particular, we have found that the number of fields and their masses are correlated for the heavy mass limit, whereas relations among the particles details and the number of fields take place at the other regimes. If after further scrutiny the LHC data turns out to be correct, our result could still allow for BSM physics, however tightly constrained.

\acknowledgments{
The authors are grateful to the National Council for Scientific and Technological Development (CNPq) and the Brazilian Federal Agency for Support and Evaluation of Graduate Education (CAPES) for financial support. IK is funded by the grant no. 303283/2022-0 and 401567/2023-0, CNPq.
}

\appendix
\section{Functional measure from operator formalism}
\label{app}

It should be noted that the equivalence between path integral and canonical quantization holds only at the formal level. It indeed depends on many choices, such as how one takes the continuum limit or the operator ordering after skeletonization (the usual choice being the Weyl ordering). Moreover, such formal equivalence has only been proven for standard theories with quadratic kinetic terms in flat backgrounds. It is much less clear whether this equivalence holds for higher-derivative theories, despite its importance in physics (and particularly in effective field theory), or in curved spaces, let alone when the configuration-space geometry is non-trivial. We take the path integral formalism as defining the quantum field theory, in the same way one constructs an effective field theory in the Wilsonian sense. Until a general proof for the equivalence between the canonical and path integral quantization is found, these formalisms should be treated as independent models of reality.
Putting all these technical issues aside, we will show below that an arbitrary choice of the configuration-space measure is consistent with the canonical quantization.

The construction of the path integral from canonical quantization essentially involves: (i) splitting up the amplitude $\langle{\Phi_f(\bold x), t_f | \Phi_0(\bold x), t_0}\rangle$ into $N$ parts, (ii) inserting completeness relations over momentum/field variables and (iii) taking the limit $N\to\infty$. 
The completeness relations would in general read~\footnote{Most textbooks construct the quantum-mechanical path integral and then promote the result to quantum field theory by simply relabelling the variables. A rather detailed construction of the path integral directly in quantum field theory can be found in Ref.~\cite{Hatfield:2019sox}.}:
\begin{align}
\int \mathcal{D} \Phi^I(\bold x) \, \ket{\Phi^I(\bold x), t}\bra{\Phi^I(\bold x), t} 
&=
\mathbb{1}
\ ,
\\
\int \mathcal{D} \Pi_I(\bold x) \, \ket{\Pi_I(\bold x), t}\bra{\Pi_I(\bold x), t}
&=
\mathbb{1}
\ ,
\end{align}
with
\begin{align}
	\mathcal{D} \Phi^I(\bold x)
	=
	\prod_{{\bold x}, I} d \Phi^I(\bold x) \mathcal{M}_\Phi(\Phi)
	\ ,
	\\
	\mathcal{D} \Pi_I(\bold x)
	=
	\prod_{{\bold x}, I} d \Pi^I(\bold x) \mathcal{M}_\Pi(\Pi)
	\ ,
\end{align}
where the product runs over the spatial coordinates.
Here $\mathcal{M}_\Phi(\Phi)$ and $\mathcal{M}_\Pi(\Pi)$ are the induced measures on the corresponding submanifolds, reflecting a non-trivial phase-space geometry/topology, and $\ket{\Phi^I(\bold x), t}, \ket{\Pi_I(\bold x), t}$ are the eigenvectors of the field and momentum operators $\hat \Phi^i$ and $\hat \Pi_i$, respectively.
Following the usual procedure and taking the continuum limit gives:
\begin{equation}
\mathcal{D}\Pi_i \mathcal{D}\Phi^i
=
\prod_{x,I} \mathcal{M}(\Pi,\Phi) \, d\Pi_I(x) d\Phi^I(x)
\ ,
\label{measure}
\end{equation}
where now the product goes over spacetime points and $\mathcal{M}(\Pi,\Phi) =  \mathcal{M}_\Phi(\Phi) \mathcal{M}_\Pi(\Pi)$.
Any phase-space measure of the form~\eqref{measure} would thus be consistent with canonical quantization.
Here $\mathcal{M}(\Pi,\Phi)$ is a generally non-trivial factor that accounts for curved geometries, non-trivial topologies, constraints (e.g. gauge choices) etc. In quantum mechanics, for example, the phase space of a spin system is a sphere, hence $\mathcal{M}(\Pi,\Phi)$ is non-trivial in this case.

By means of Darboux's theorem, one could always set $\mathcal{M}(\Pi,\Phi) = 1$ locally. However, the path integral is a global object, receiving contributions of field configurations from afar. Enforcing $\mathcal{M}(\Pi,\Phi) = 1$ in the path integral for a non-trivial topology would thus require many different phase-space charts in order to compute the path integral, which is not how the calculation is usually performed. It is a lot easier just to keep the non-trivial measure during the integration.

The standard approach (where one identifies the metric from the kinetic term) assumes that the phase-space measure is globally trivial, i.e. $\mathcal{M}(\Pi,\Phi) \equiv 1$. But already in classical and quantum mechanics, one finds examples where this cannot be true. Thus it seems that the standard approach misses out on interesting phenomena.

In the standard approach, one considers theories with flat phase space and without higher derivatives, namely:
\[
S
=
\int
\mathrm{d}^4x \,
G_{IJ}\partial_\mu \Phi^I \partial^\mu \Phi^J + \cdots
\ .
\]
In this case, one can then integrate the momentum variable to wit:
\[
Z
=
\int \prod_{x}d\Phi^I (\text{det}\, G_{IJ})^{1/2}\, e^{i S[\Phi]}
\ .
\]
Therefore, the configuration-space measure acquires a factor of the kinetic term coefficient $G_{IJ}$. From the theory of integrations on manifolds, one readily sees that $G_{IJ}$ plays the role of a metric, thus justifying the widespread definition of the kinetic term coefficient $G_{IJ}$ as the configuration-space metric.

We stress that the connection between the kinetic term and the configuration-space metric relies on the following assumptions: (i) an unconstrained phase space with trivial geometry and topology and (ii) the absence of higher derivatives. However, like we mentioned before, there are important examples in physics that do not satisfy these hypotheses.
A curved phase space would require a non-trivial factor $\mathcal{M}(\Pi,\Phi) \neq 1$, in which case the momentum integral would not be Gaussian. After solving the momentum integral, the configuration-space measure would receive other corrections. The comparison with the Riemannian measure would no longer allow for the interpretation of $G_{IJ}$ as the metric. For example, if  $\mathcal{M}(\Pi,\Phi) =  \mathcal{M}(\Phi)$ does not depend on $\Pi_i$, the configuration-space measure would read:
\[
\prod_{x}d\Phi^I \mathcal{M}(\Phi) (\text{det}\, G_{IJ})^{1/2}
\ .
\]
The true configuration-space metric $\mathfrak g_{IJ}$ would rather be:
\[
\text{det} \, \mathfrak g_{IJ}
=
\mathcal{M}(\Phi)^2 \, \text{det}\, G_{IJ}
\ .
\]
The metric $\mathfrak g_{IJ}$ is perfectly consistent with canonical quantization and cannot be identified from the kinetic term. Our construction of the configuration-space measure as an expansion in inverse powers of the cutoff is an effective way of accounting for $\mathcal{M}(\Phi)$ (or more generally $\mathcal{M}(\Pi,\Phi)$).

%
%
%
%

%
\end{document}